\documentclass[12pt]{iopart}
\usepackage{amssymb}
\usepackage{color}
\usepackage{graphicx,subfigure}
\usepackage{bm}
\def\comment#1{}

\def\slashchar#1{\setbox0=\hbox{$#1$}           
   \dimen0=\wd0                                 
   \setbox1=\hbox{/} \dimen1=\wd1               
   \ifdim\dimen0>\dimen1                        
      \rlap{\hbox to \dimen0{\hfil/\hfil}}      
      #1                                        
   \else                                        
      \rlap{\hbox to \dimen1{\hfil$#1$\hfil}}   
      /                                         
   \fi}                                         %

\begin{document}

\title{Strong-coupling topological Josephson effect in quantum wires}

\author{Flavio S. Nogueira and Ilya Eremin}

\address{Institut f{\"u}r Theoretische Physik III, Ruhr-Universit\"at Bochum,
Universit\"atsstra\ss e 150, 44801 Bochum, Germany
}
\pacs{74.50.+r, 74.78.Na, 73.63.Nm}

\begin{abstract}
We investigate the Josephson effect for a setup with
two lattice quantum wires featuring Majorana zero energy boundary modes at the tunnel junction.
In the weak-coupling, the exact solution reproduces the perturbative result for the energy containing a contribution $\sim \pm\cos(\phi/2)$ relative to the
tunneling of paired Majorana fermions. As the tunnel amplitude $g$ grows relative
to the hopping amplitude $w$, the gap between the energy levels gradually diminishes until it closes completely
at the critical value $g_c=\sqrt{2}w$. At this point the Josephson energies
have the principal values
$E_{m\sigma}=2\sigma\sqrt{2}w\cos[\phi/6+2\pi (m-1)/3]$, where
$m=-1,0,1$ and $\sigma=\pm 1$, a result not following from perturbation theory.
It represents a transparent regime where three Bogoliubov states merge, leading to additional degeneracies of the topologically nontrivial
ground state with odd number of Majorana fermions at the end of each wire. We also obtain the exact tunnel currents for a fixed parity of the eigenstates.
The Josephson current shows the characteristic $4\pi$ periodicity expected for a topological Josephson effect. We discuss the additional features
of the current associated with a closure of the energy gap between the energy levels.
\end{abstract}

\maketitle

\section{Introduction}

In the Josephson effect \cite{Tinkham} the phase coherent tunneling across a junction between two systems A and B with broken $U(1)$ symmetry
implies a dissipationless current oscillating with $\phi\equiv\phi_a-\phi_b$, where $\phi_a$ and $\phi_b$ denote
the phases of the superfluid order parameter in A and B, respectively. The typical oscillating behavior in a conventional Josephson effect is
given by  $I=I_0\sin\phi$, having periodicity $2\pi$. Recently the Josephson effect has been considered
in the framework of topological insulators \cite{Fu-Kane-2009}. It was shown that in a system where the tunnel
junction is a topological insulator, a fractional Josephson current $I\propto\pm\sin(\phi/2)$ occurs as a consequence of the $Z_2$ topological nature of
the quantum spin Hall insulator. The occurrence of a contribution in the current featuring half of the phase difference
has in this case a topological origin. In particular, the $4\pi$ periodicity arises due to boundary zero 
fermionic modes at the junction. Since these boundary
modes have zero energy, they are not
influenced by changes in the magnitude of the gap in the bulk superconducting state \cite{Fu-Kane-2009,Kitaev}.  Interestingly, these zero energy
boundary modes are found to be Majorana fermions, which in this context correspond to Bogoliubov quasi-particles having
the reality property $\gamma=\gamma^\dagger$. In order to exhibit  a concrete example of Bogoliubov quasi-particles fulfilling
the Majorana reality condition, it is enough to recall the Bogoliubov transformation in superconductors \cite{Tinkham},
\begin{equation}
\label{Eq:Bogs}
 \gamma_{1{\bf k}}=u_{\bf k}c_{{\bf k},\alpha}-v_{\bf k}c_{-{\bf k},\beta}^\dagger, ~~~~~~~~\gamma_{2{\bf k}}=u_{\bf k}c_{-{\bf k},\alpha}
+v_{\bf k}c_{{\bf k},\beta}^\dagger,
\end{equation}
where $|u_{\bf k}|^2+|v_{\bf k}|^2=1$. For conventional spin singlet superconductors we have $\alpha=\uparrow$ and $\beta=\downarrow$, in which case
the reality condition cannot be achieved. Thus, let us consider $\alpha=\beta$ or, more simply, spinless fermions. If
$u_{\bf k}=1/\sqrt{2}$ and $v_{\bf k}=e^{i\zeta_{\bf k}}/\sqrt{2}$, we will have that at zero energy
(i.e., $k=0$) $\gamma_{10}$ will be a Majorana fermion
if $\zeta_0=\pi/2$, while $\gamma_{20}$ will be a Majorana fermion if $\zeta_0=0$. 
In the specific model considered by Fu and Kane \cite{Fu-Kane-2009,Fu-Kane-2008}, superconductivity was induced by proximity
effect on the surface of a topological insulator, with surface excitations being described by a Dirac-like Hamiltonian in two dimensions.
Within a mean-field treatment of the problem, the $U(1)$ symmetry breaking is introduced by coupling the anomalous fermion bilinears
$\psi_\uparrow^\dagger\psi_\downarrow^\dagger$ and $\psi_\downarrow\psi_\uparrow$ to an uniform superconducting gap
$\Delta=\Delta_0e^{i\phi}$. Thus, one can define spinless fermionic fields entering Eq. (\ref{Eq:Bogs}) as
$c_{\bf k}=(\psi_{{\bf k},\uparrow}+e^{i\theta_{\bf k}}\psi_{{\bf k},\downarrow})/\sqrt{2}$ \cite{Fu-Kane-2008}. For the case of
two superconducting surfaces connected via a nanowire, one can assume the system to be one-dimensional and compute the corresponding
fractional Josephson current \cite{Fu-Kane-2009,Fu-Kane-2008,Sarma-2010}.

As another instance of the theory considered in Refs. \cite{Fu-Kane-2009} and \cite{Fu-Kane-2008}, it
is interesting to notice that beyond mean-field theory its topological content in a nanowire setting is similar to the
model of Goldstone and Wilczek in 1+1 dimensions \cite{Goldstone-Wilczek-1981} involving
Dirac fermions coupled to two scalar fields $\varphi_1$ and $\varphi_2$.  Indeed, we can take $\varphi_1$ and $\varphi_2$
as the real and imaginary parts of the superconducting order parameter, $\varphi$. In this way, the expectation value of
the covariant fermionic current $j^\mu$ can be shown to be identical to the topological current \cite{Goldstone-Wilczek-1981},
\begin{equation}
 K^\mu=\langle j^\mu\rangle=\frac{1}{2\pi}\epsilon^{\mu\nu}\frac{i(\varphi^*\partial_\nu\varphi-\varphi\partial_\nu\varphi^*)}
{|\varphi|^2}.
\end{equation}
Thus, if $L$ is the length of the wire, the topological charge,
\begin{eqnarray}
 Q=\int dx^1K^0&=&\frac{1}{2\pi}\int_{-L/2}^{L/2}dx^1\frac{\partial}{\partial x^1}\arctan\left(\frac{\varphi_2}{\varphi_1}\right)
\nonumber\\
&=&\frac{1}{2\pi}\left[\left.\arctan\left(\frac{\varphi_2}{\varphi_1}\right)\right|_{L/2}
-\left.\arctan\left(\frac{\varphi_2}{\varphi_1}\right)\right|_{-L/2}\right]
\nonumber\\
&=&\frac{\Delta\phi}{2\pi},
\end{eqnarray}
yields the charge of the topological excitation (soliton) in terms of the phase difference $\Delta\phi$ between the boundaries only, and
the order parameter in the bulk plays no role. Let us assume that each end of the wire is located in superconductors
$A$ and $B$, with corresponding phases $\phi_a$ and $\phi_b$, respectively. In this case we have
$\varphi_1(-L/2)=\Delta_{0,a}\cos\phi_a(-L/2)$, $\varphi_2(-L/2)=\Delta_{0,a}\sin\phi_a(-L/2)$, $\varphi_1(L/2)=\Delta_{0,b}\cos\phi_b(L/2)$,
and $\varphi_2(L/2)=\Delta_{0,b}\sin\phi_b(L/2)$, so that
\begin{equation}
\label{Eq:dphi}
 \Delta\phi=\arctan[\tan\phi_b(L/2)]-\arctan[\tan\phi_a(-L/2)],
\end{equation}
which for both $\phi_a(-L/2)$ and $\phi_b(L/2)$
in the interval $(-\pi/2,\pi/2)$ yields the principal value $\Delta\phi=\phi_b(L/2)-\phi_a(-L/2)$.
The function $\arctan(\tan\phi)$ is double-valued at $\phi=(2k+1)\pi/2$, $k\in\mathbb{Z}$ and has principal value $\phi$
in the interval $-\pi/2<\phi<\pi/2$. Thus, as the phase $\phi$ at one of the ends of the wire vary from $0$ to $2\pi$,
the $\arctan(\tan\phi)$ makes an abrupt change of sign at $\pi/2$.
In the context of Eq. (\ref{Eq:dphi}) the phases at the boundaries are constrained to half of the period of the order parameter,
reflecting the fact that solitons (and Bogoliubov quasi-particles as well) are created in pairs, and the full period corresponds
to two solitons. Therefore, we should take $\phi$ varying from $-\pi/2$ to $\pi/2$, e. g.,
$\phi_a(-L/2)=-\pi/2$ and $\phi_b(L/2)=\pi/2$,
so that $\Delta\phi=\pi$, implying the fractional charge $Q=1/2$. It is of course also possible to choose
$\phi_a(-L/2)=\pi/2$ and $\phi_b(L/2)=-\pi/2$ and obtain $Q=-1/2$. This pair of solitons of charge 1/2 corresponds
to the Majorana boundary modes in the wire. This fractional charge of a Majorana fermion can also be directly checked
using the spinless variant of the Bogoliubov quasi-particles (\ref{Eq:Bogs}) at $k=0$. Indeed, using our
previous discussion after Eq. (\ref{Eq:Bogs}), we see that for $\zeta_0=\pi/2$ we have
a particle number operator $\gamma_{10}^\dagger\gamma_{10}=\gamma_{10}^2=1/2$, with a similar result holding for the
Bogoliubov quasi-particle $\gamma_{20}$ when $\zeta_0=0$.

%

All the physical and mathematical properties discussed above are also present in a simple one-dimensional lattice
model introduced by Kitaev some time ago \cite{Kitaev}.
This model has recently been used to motivate interesting realizations of physical systems where Majorana fermions
may play a crucial role
\cite{Sarma-2010,Sengupta-2003,Felix-2010,Alicea-2010,Feigelman-2010,Loss-2011,Tanaka-2009,sh-2010,Neupert-2010,Asano-2010,Linder-2010}.
Within the framework of Kitaev's model, we will investigate two chains of spinless fermions having broken $U(1)$ symmetry connected by a tunnel junction.
We will first assume that the broken symmetry state is such that the superconducting gap $|\Delta|=w$, where $w$ is the hopping between nearest neighbor sites. This
simplifying assumption has the advantage of allowing an exact derivation of the energy spectrum of the Josephson effect. The exact solution unveils
interesting features at strong-coupling which may possibly be present in more realistic situations involving semiconducting wires having strong
spin-orbit coupling lying
on the surface of an s-wave superconductor \cite{Felix-2010,Alicea-2010}.
One salient feature at strong Josephson coupling is the existence of a critical value of the tunnel coupling, $g_c=\sqrt{2}w$,
leading to level crossing between all the energy levels having different quantum numbers $m$, thus closing the gap between
the levels and increasing the
degeneracy.
For such a value of the coupling the energy levels can be equivalently written as crossing levels of
the form
$E_{m\sigma}(\phi)=2\sigma\sqrt{2}w\cos(\phi/6-2\pi (m-1)/3)$, where $m=-1,0,1$ and $\sigma=\pm 1$, quite different from the leading order weak-coupling
result featuring an energy $\sim \cos(\phi/2)$ \cite{Fu-Kane-2009,Felix-2010,Alicea-2010}. In order to check whether our results are generic, we
also consider the numerical solution of the problem when $|\Delta|\neq w$ and show that the same gap closing feature is present in
the more general situation, although the structure of the spectrum is more complex. Despite the $\phi/6$ factor in the
argument of the cosine, the Josephson current still exhibits the $4\pi$ periodicity characteristic of the topological Josephson effect.
In fact, we will show that the factor $\phi/6$ actually corresponds to the principal value of a certain double-valued function
$\theta(\phi)$ having $4\pi$ periodicity.
At the same time, the current will be shown to exhibit some additional features which arise due to closure of the energy gap.

The plan of the paper is as follows. In Section 2 we introduce the model and obtain the exact energy levels for the
Josephson effect. Both situations corresponding to $|\Delta|=w$ and  $|\Delta|\neq w$ are discussed. The former
case will allow us to write down simple exact analytical expressions for the energy eigenvalues, while the latter case
can be diagonalized exactly.
In Section 3 we calculate the Josephson current from both equilibrium statistical mechanics and
using an ensemble where the states have a fixed parity \cite{Tuominen-1992}.  Section 4 concludes the paper.

\section{The Kitaev model based Josephson junction}

The Hamiltonian of the system consists of two wires $A$ and $B$ which are in contact via a tunneling Hamiltonian
$H=H_A+H_B+H_T$
\cite{Kitaev,Alicea-2010},
%
where
\begin{equation}
 H_A=\sum_{j=1}^{N-1}[-w(a_j^\dagger a_{j+1}+a_{j+1}^\dagger a_j)+\Delta_a a_{j+1}^\dagger a_j^\dagger+
\Delta_a^* a_j a_{j+1}],
\end{equation}
\begin{equation}
 H_B=\sum_{j=1}^{N-1}[-w(b_j^\dagger b_{j+1}+b_{j+1}^\dagger b_j)+\Delta_b b_{j+1}^\dagger b_j^\dagger+
\Delta_b^* b_j b_{j+1}],
\end{equation}
%
\begin{equation}
 H_\Gamma=g( a_N^\dagger b_1+ b_1^\dagger a_N),
\end{equation}
where $\Delta_a=|\Delta|e^{i\phi_a}$ and $\Delta_b=|\Delta|e^{i\phi_b}$, with the magnitudes of the superconducting gaps assumed
to be the same. If we perform the global gauge transformations $a_j\to e^{i\phi_a/2}a_j$ and  $b_j\to e^{i\phi_b/2}b_j$,
the phases $\phi_a$ and $\phi_b$ are gauged away in the Hamiltonians $H_A$ and $H_B$, while the tunnel Hamiltonian becomes,
\begin{equation}
 H_T=\Gamma a_N^\dagger b_1+ \Gamma^*b_1^\dagger a_N,
\end{equation}
where $\Gamma=ge^{\phi/2}$, with $\phi\equiv\phi_b-\phi_a$ being the phase difference across the junction.

We will first consider the case where $|\Delta|=w$, where an exact analytic solution for the Josephson energy eigenstates is possible.

It is convenient to write the Hamiltonians $H_A$ and $H_B$ in matrix form. Thus, we will write $H_A=(1/2)\psi_A^\dagger M \psi_A$, where
$\psi_A^\dagger=[a_1^\dagger~~a_2^\dagger~\cdots~a_N^\dagger~~a_1~~a_2~\cdots~a_N]$ and $M$ is a symmetric $2N\times 2N$ matrix having zero trace. The
Hamiltonian $H_B$ is written similarly with the $a_j$ operators being replaced by the $b_j$ ones. Note that
$\psi_A^\dagger\psi_A=N$.
The matrix $M$ has a doubly degenerated zero energy eigenvalue corresponding to two Majorana zero energy modes residing
on each end of the chain, and $(N-1)$-fold
degenerated energy eigenvalues $\pm 2w$. The Hamiltonian can be written in diagonal form by introducing the following fermionic operators,
\begin{equation}
\label{Majos}
 \gamma_{1}=\frac{1}{\sqrt{2}}(a_1+a_1^\dagger),~~~~~~~~\gamma_{N}=\frac{i}{\sqrt{2}}(a_N^\dagger-a_N),
\end{equation}
%
which are the zero Majorana boundary modes, and the nonzero modes,
%
\begin{eqnarray}
\label{subL1}
&c_{N-(2n+1),\sigma}=\frac{1}{\sqrt{2n+1}}\left[\frac{1}{2}(-\sigma a_1+\sigma a_1^\dagger+a_{2n+2}+a_{2n+2}^\dagger)
\right.\nonumber\\
&+\left.(1-\delta_{n0})\sum_{m=1}^n(a_{2m}-\sigma a_{2m+1})\right],
\end{eqnarray}
\begin{eqnarray}
\label{subL2}
  c_{N-2(k+1),\sigma}&=&\frac{1-\delta_{N2}}{\sqrt{2(k+1)}}\left[\frac{1}{2}(a_1- a_1^\dagger+a_{2k+1}+a_{2k+1}^\dagger)
\right.\nonumber\\
&+&\left.\sum_{m=0}^k(-\sigma)^{m+1} a_{m+2}\right],
\end{eqnarray}
where $\sigma=\pm 1$, $n=0,1,\dots,n_N$, and $k=0,1,\dots,k_N$. Here $n_N=N/2-1$ and $k_N=N/2-2$ if $N$ is even, and $n_N=(N-1)/2$ and $k_N=(N-3)/2$ if $N$ is odd.
The operators (\ref{subL1}) and (\ref{subL2}) correspond to two interpenetrating sublattices $L_1$ and $L_2$.
The fermionic operators satisfy the local constraint,
\begin{equation}
\label{constraint}
 \sum_{\sigma=\pm 1}c_{j\sigma}^\dagger c_{j\sigma}=1,
\end{equation}
which together with the constraint
\begin{equation}
 \gamma_{1}^\dagger \gamma_{1}=\gamma_1^2=\gamma_{N}^\dagger \gamma_{N}=\gamma_N^2=1/2,
\end{equation}
for the zero modes yields
\begin{equation}
 \gamma_{1}^\dagger \gamma_{1}+\gamma_{N}^\dagger \gamma_{N}+\sum_{j=1}^{N-1}\sum_{\sigma=\pm 1}c_{j\sigma}^\dagger c_{j\sigma}=N.
\end{equation}
Therefore, the Hamiltonian for a chain A with $N$ sites has the diagonal form,
\begin{equation}
\label{Ha}
 H_A=w\sum_{j=1}^{N-1}\sum_{\sigma=\pm 1}\sigma c_{j\sigma}^\dagger c_{j\sigma}.
\end{equation}
In view of the constraint (\ref{constraint}) the above representation of the Hamiltonian $H_A$ corresponds to localized spins $1/2$ in an
external magnetic field of magnitude $2w$ applied along the $z$-direction. Note that we
can use the constraint (\ref{constraint}) in Eq. (\ref{Ha}) to rewrite it in the form
\begin{equation}
 H_A=2w\sum_{j=1}^{N-1}(c_{j,+1}^\dagger c_{j,+1}-1/2),
\end{equation}
obtained in Ref. \cite{Kitaev}.

The Hamiltonian $H_B$ is diagonalized in a similar fashion, except that we have to name the new fermionic operators differently, say $d_{j\sigma}$
for the nonzero modes, with $\delta_{1}$ and $\delta_{N}$ being the zero boundary modes.  Since in the new basis the
Hamiltonians $H_A$ and $H_B$ are written in terms of localized particles, it follows that the part of the spectrum corresponding to the Josephson energy
levels can be obtained by solving a reduced Hamiltonian, basically one featuring two dimers ($H_A$ and $H_B$ for $N=2$)
connected by a tunnel junction. This procedure is basically the same as the one
used by Alicea {\it et al.} \cite{Alicea-2010} to solve the problem perturbatively.

For the exact result we obtain a spectrum containing two zero energy eigenvalues corresponding to
Majorana boundary modes, the energy levels $\pm 2w$, each with degeneracy  $2(N-2)$, and
the phase-dependent Josephson energies obtained by solving the dimer/dimer junction, whose Hamiltonian is
the same as $H$ for $N=2$, i.e.,
\begin{equation}
 H=\frac{1}{2}\psi^\dagger M\psi,
\end{equation}
where
\begin{equation}
\psi^\dagger=\left[
\begin{array}{cccccccc}
a_1^\dagger & a_2^\dagger & a_1 & a_2 & b_1^\dagger & b_2^\dagger & b_1 & b_2
\end{array}
\right],~~~~
\psi=
\left[
\begin{array}{c}
a_1\\
\noalign{\medskip}
a_2\\
\noalign{\medskip}
a_1^\dagger\\
\noalign{\medskip}
a_2^\dagger\\
\noalign{\medskip}
b_1\\
\noalign{\medskip}
b_2\\
\noalign{\medskip}
b_1^\dagger\\
\noalign{\medskip}
b_2^\dagger
\end{array}
\right],
\end{equation}
\begin{equation}
\label{matrix}
M=\left[
\begin{array}{cccccccc}
 0 & -w & 0 & -w & 0 & 0 & 0 & 0\\
\noalign{\medskip}
-w & 0 & w & 0 & \Gamma & 0 & 0 & 0\\
\noalign{\medskip}
0 & w & 0 & w & 0 & 0 & 0 & 0\\
\noalign{\medskip}
-w & 0 & w & 0 & 0 & 0 & -\Gamma^* & 0\\
\noalign{\medskip}
0 & \Gamma^* & 0 & 0 & 0 & -w & 0 & -w\\
\noalign{\medskip}
0 & 0 & 0 & 0 & -w & 0 & w & 0\\
\noalign{\medskip}
0 & 0 & 0 & -\Gamma & 0 & w & 0 & w\\
\noalign{\medskip}
0 & 0 & 0 & 0 & -w & 0 & w & 0
 \end{array}
\right]
\end{equation}
The above dimer/dimer problem is exactly diagonalizable by a Bogoliubov transformation.
Before proceeding with this calculation, let us first diagonalize each dimer separately in order to
gain some physical insight from the terms arising in the tunnel Hamiltonian. The dimer Hamiltonian for the first
chain is diagonalized by the transformation,
\begin{equation}
 \gamma_1=\frac{1}{\sqrt{2}}(a_1+a_1^\dagger),
\end{equation}
\begin{equation}
 \gamma_2=\frac{i}{\sqrt{2}}(a_2^\dagger-a_2),
\end{equation}
\begin{equation}
 c_-=\frac{1}{2}(a_1-a_1^\dagger+a_2+a_2^\dagger),
\end{equation}
\begin{equation}
 c_+=\frac{1}{2}(-a_1+a_1^\dagger+a_2+a_2^\dagger),
\end{equation}
where $\gamma_1$ and $\gamma_2$ are Majorana boundary modes satisfying $\gamma_1^2=\gamma_2^2=1/2$,
and the constraint
\begin{equation}
\label{constraint-dimer}
 c_+^\dagger c_++c_-^\dagger c_-=1
\end{equation}
holds.
Note that the equations above are just the general transformation given in Eqs. (\ref{Majos},\ref{subL1},\ref{subL2}) for the
special case $N=2$.
Thus, the Hamiltonian for the dimer A is in this new operator basis given by,
\begin{equation}
 H_A=w(c_+^\dagger c_+-c_-^\dagger c_-)=2w\left(c_+^\dagger c_+-\frac{1}{2}\right).
\end{equation}
Note that the Majorana fermions do not appear in the Hamiltonian $H_A$, since their energy eigenvalues vanish. We obtain a similar
result for the Hamiltonian $H_B$, with the new fermionic operators being called $\delta_1$, $\delta_2$,  $d_+$, and $d_-$. Here
$\delta_1$ and $\delta_2$ are the corresponding Majorana boundary modes for the dimer Hamiltonian $H_B$.

The Majorana fermion operators will appear explicitly only in the tunnel Hamiltonian connecting the states of the two dimers. This
will make two of the four Majorana modes overlap, making in this way two zero modes disappear. This is better seen
by collecting the Majorana fermions at the junction into a new (ordinary) fermionic operator defined by,
\begin{equation}
 f=\frac{1}{\sqrt{2}}(\gamma_2+i\delta_1).
\end{equation}
Thus, the Hamiltonian of the dimer/dimer system can be written in the form,
\begin{equation}
\label{H-dd}
 H=H_0+\widetilde H,
\end{equation}
where
\begin{equation}
 H_0=w(c_+^\dagger c_+-c_-^\dagger c_-+d_+^\dagger d_+-d_-^\dagger d_-)-g\cos\left(\frac{\phi}{2}\right)\left(f^\dagger f-\frac{1}{2}\right),
\end{equation}
and
\begin{eqnarray}
 \widetilde H&=&\frac{ge^{i\phi/2}}{4}[i(c_+^\dagger+c_-^\dagger)(f^\dagger-f)-i(f+f^\dagger)(d_--d_+)
\nonumber\\
&+&c_+^\dagger d_-+c_-^\dagger d_--c_+^\dagger d_+-c_-^\dagger d_+]+{\rm h.c.}
\end{eqnarray}
The Hamiltonian $\widetilde H$ describes
the hybridization between the Bogoliubov quasi-particles from
$H_0$ and, in addition, tunneling processes involving  fused Majorana states
across the junction.

Let us first neglect the hybridization contribution and study the spectrum of the Hamiltonian $H_0$. The energy eigenstates are
given by $|n_+^c,n_-^c;n_+^d,n_-^d;n_f\rangle$, which are also eigenstates of the particle number operators
$N_{\pm}^c=c_\pm^\dagger c_\pm$, $N_\pm^d=d_\pm^\dagger d_\pm$, and $N_f=f^\dagger f$, since these operators commute with
$H_0$. In view of the constraint (\ref{constraint-dimer}), there are also corresponding constraints for the quantum numbers, i.e.,
\begin{equation}
\label{constraint-ns}
n_+^c+n_-^c=1, ~~~~~~~~n_+^d+n_-^d=1.
\end{equation}
In Table 1 we show the eigenstates of $H_0$ and the respective energy eigenvalues. We note that
the eigenstates $|1,0;0,1;0\rangle$ and $|0,1;1,0;0\rangle$ are twofold degenerate
with energy eigenvalues $E_{10010}=E_{01100}=(g/2)\cos(\phi/2)$. A twofold degeneracy
also occurs with the eigenstates $|1,0;0,1;1\rangle$ and $|0,1;1,0;1\rangle$, which have
energy eigenvalues $E_{10011}=E_{01101}=-(g/2)\cos(\phi/2)$. These degenerate energies vanish for $\phi=(2k+1)\pi$, with
$k\in\mathbb{Z}$. A closer look at Table 1 shows that we can attribute either a positive or negative sign to the coefficient
of $\cos(\phi/2)$, depending on whether or not the occupation number $n_f$ vanish. The coefficient of $2w$ is either $+1$, $-1$, or
$0$, depending on the configuration involving the occupation numbers $n_\pm^c$ and $n_\pm^d$. Thus, we see that the general
expression for the energy eigenvalues of $H_0$ can be written as,
\begin{equation}
 E_{m\sigma}^{(0)}=2mw+\frac{\sigma g}{2}\cos\left(\frac{\phi}{2}\right),
\end{equation}
where
\begin{equation}
 m=\frac{n_+^c+n_+^d-n_-^c-n_-^d}{2}, ~~~~~~~~~\sigma=(-1)^{n_f}
\end{equation}
with the understanding that the constraints (\ref{constraint-ns}) have to be satisfied, and $n_f=0,1$.
Thus, the quantum number $m$ is physically the $z$-projection of the {\it total} pseudospin ``magnetic'' quantum
number. The quantum number
$\sigma=(-1)^{n_f}$ is the parity of the fused Majorana state.

\begin{table}
\begin{center}
\caption{Eigenstates $|n_+^c,n_-^c;n_+^d,n_-^d;n_f\rangle$ of $H_0$ and their energy eigenvalues. The
last slot in the eigenstates denote the presence or absence of a fused Majorana fermion.}
\begin{tabular}{ | l |l |}
\hline
Eigenstate & Energy eigenvalue\\ \hline \hline
$|1,0;1,0;0\rangle$ & $E_{10100}=2w+\frac{g}{2}\cos\left(\frac{\phi}{2}\right)$ \\ \hline
$|0,1;0,1;0\rangle$ & $E_{01010}=-2w+\frac{g}{2}\cos\left(\frac{\phi}{2}\right)$\\ \hline
$|1,0;0,1;0\rangle$ & $E_{10010}=\frac{g}{2}\cos\left(\frac{\phi}{2}\right)$ \\ \hline
$|0,1;1,0;0\rangle$ & $E_{01100}=\frac{g}{2}\cos\left(\frac{\phi}{2}\right)$ \\ \hline
$|1,0;1,0;1\rangle$ & $E_{10101}=2w-\frac{g}{2}\cos\left(\frac{\phi}{2}\right)$ \\ \hline
$|0,1;0,1;1\rangle$ & $E_{01011}=-2w-\frac{g}{2}\cos\left(\frac{\phi}{2}\right)$ \\ \hline
$|1,0;0,1;1\rangle$ & $E_{10011}=-\frac{g}{2}\cos\left(\frac{\phi}{2}\right)$\\ \hline
$|0,1;1,0;1\rangle$ & $E_{01101}=-\frac{g}{2}\cos\left(\frac{\phi}{2}\right)$\\ \hline
\end{tabular}
\end{center}
\end{table}

The exact energy spectrum is easily obtained from the diagonalization of the matrix (\ref{matrix}), which leads to
the  secular equation,
\begin{equation}
\label{cubic}
 E^3-(4w^2+g^2)E\pm gw^2\cos(\phi/2)=0,
\end{equation}
and the Hamiltonian of the effective dimer/dimer system describing the tunnel junction becomes,
\begin{equation}
\label{Eq:dimer}
 H_{\rm Junction}=\frac{1}{2}\sum_{m,\sigma}E_{m\sigma}N_{m\sigma},
\end{equation}
where $N_{m\sigma}$ are the corresponding particle number operators.
Therefore, we obtain along with a doubly degenerated zero energy eigenvalue, the six energy eigenvalues,
\begin{eqnarray}
\label{energies}
 E_{m\sigma}(\phi)&=&(2gw^2)^{1/3}\left\{e^{-2i\pi (m-1)/3}\left[\sigma\cos\left(\frac{\phi}{2}\right)-{\rm i}G(\phi)
\right]^{1/3}
\right.\nonumber\\
&+&\left.e^{2i\pi (m-1)/3}\left[\sigma\cos\left(\frac{\phi}{2}\right)+{\rm i}G(\phi)
\right]^{1/3}\right\},
\end{eqnarray}
where $m=-1,0,1$ and $\sigma=\pm 1$ as before, and
\begin{equation}
\label{G}
 G(\phi)=\sqrt{\frac{1}{4g^2w^4}\left(\frac{g^2+4w^2}{3}\right)^3-\cos^2\left(\frac{\phi}{2}\right)}.
\end{equation}
The following relation holds,
\begin{equation}
\label{Eq:energ-parity}
 E_{m,\sigma}(\phi)=-E_{-m,-\sigma}(\phi),
\end{equation}
along with the constraint, 
\begin{equation}
\label{Eq:constraint-junction}
 N_{m\sigma}+N_{-m,-\sigma}=1,
\end{equation}
as expected for Bogoliubov quasi-particles. This allows us to rewrite the effective dimer/dimer Hamiltonian 
(\ref{Eq:dimer}) of the junction as,
\begin{equation}
\label{Eq:H-Junction}
 H_{\rm Junction}=\sum_{i=1}^3E_i\left(N_i-\frac{1}{2}\right),
\end{equation}
where $E_1\equiv E_{0,-1}$, $E_2\equiv E_{1,+1}$, and
$E_3\equiv E_{1,-1}$, with a similar relabeling for the particle number operators. The form (\ref{Eq:H-Junction}) will be 
useful in the calculation of the tunnel current later.   

The energy eigenvalues above have subtle analytic properties. In fact, in view of the property (\ref{Eq:energ-parity}) and the oscillatory behavior
of the energies, we can see that pairs of energy eigenvalues having the same $m$ and opposite $\sigma$ necessarily cross at $\phi=(2k+1)\pi$,
$k\in\mathbb{Z}$. Thus, the $4\pi$ periodicity of the energies (\ref{energies}) is in this case equivalent to $2\pi$-periodicity with
double-valuedness at $\phi=(2k+1)\pi$, a fact already anticipated by Kitaev \cite{Kitaev}. Therefore, we can rewrite the energies (\ref{energies})
in terms of its principal values in the interval $-\pi<\phi<\pi$ as,
\begin{equation}
 \label{Eq:energies-2-valued}
E_{m\sigma}(\phi)=2\sigma(2gw^2)^{1/3}\cos\left[\theta(\phi)+\frac{2\pi(m-1)}{3}\right],
\end{equation}
where
\begin{equation}
\label{theta}
 \theta(\phi)=\frac{1}{3}\arctan\left[\frac{G(\phi)}{\cos(\phi/2)}\right].
\end{equation}

We emphasize that the cubic equations determining the exact Josephson energy levels always arise in the case where $|\Delta|=w$, regardless
of the size of the chains involved. For chains $A$ and $B$ having each $N$ lattice sites we obtain the characteristic
polynomial for the corresponding $2N\times 2N$ matrix $M$,
\begin{eqnarray}
 P(E)&=&E^2(E^2-4w^2)^{2N-4}[E^3-(4w^2+g^2)E-gw^2\cos(\phi/2)]
\nonumber\\
&\times&[E^3-(4w^2+g^2)E+gw^2\cos(\phi/2)].
\end{eqnarray}
Therefore, the Josephson energies are independent of the size of the wires involved, reflecting the topological nature of the
Josephson effect in such a system.

For $g\ll w$ we obtain up to second order in $g/w$,
\begin{eqnarray}
 E_{m\sigma}&\approx&\frac{4w}{\sqrt{3}}\sin\left(\frac{2\pi m}{3}\right)-\sigma g\cos\left(\frac{2\pi m}{3}\right)\cos\left(\frac{\phi}{2}\right)
\nonumber\\
&-&\frac{g^2}{16\sqrt{3}}\sin\left(\frac{2\pi m}{3}\right)(5-3\cos\phi),
\end{eqnarray}
which yields,
\begin{equation}
\label{approx0}
 E_{+1,\sigma}(\phi)\approx 2w+\frac{\sigma g}{2}\cos\left(\frac{\phi}{2}\right)+\frac{g^2}{32w}(5-3\cos\phi),
\end{equation}
\begin{equation}
\label{approx1}
E_{-1,\sigma}(\phi)\approx -2w+\frac{\sigma g}{2}\cos\left(\frac{\phi}{2}\right)-\frac{g^2}{32w}(5-3\cos\phi),
\end{equation}
and
\begin{equation}
\label{approx2}
 E_{0,\sigma}(\phi)\approx-\sigma g\cos\left(\frac{\phi}{2}\right),
\end{equation}
agreeing with the perturbative result by Alicea {\it et al.} \cite{Alicea-2010}. From the perturbative
expansion it is seen how the degeneracy of the $m=0$ eigenstate of $H_0$ is lifted when $\phi\neq (2k+1)\pi$,
$k\in\mathbb{Z}$. Note that while the first two terms in Eqs. (\ref{approx0}) and (\ref{approx1}) agree with
the energy eigenvalues $E_{10100}$, $E_{01010}$, $E_{10101}$, and $E_{01011}$ of $H_0$, Eq. (\ref{approx2})
is twice the energies $E_{10010}$, $E_{01100}$, $E_{10011}$, and $E_{01100}$ (see Table 1).

\begin{figure}
\begin{center}
\hspace{-0.5cm} \includegraphics[width=12cm]{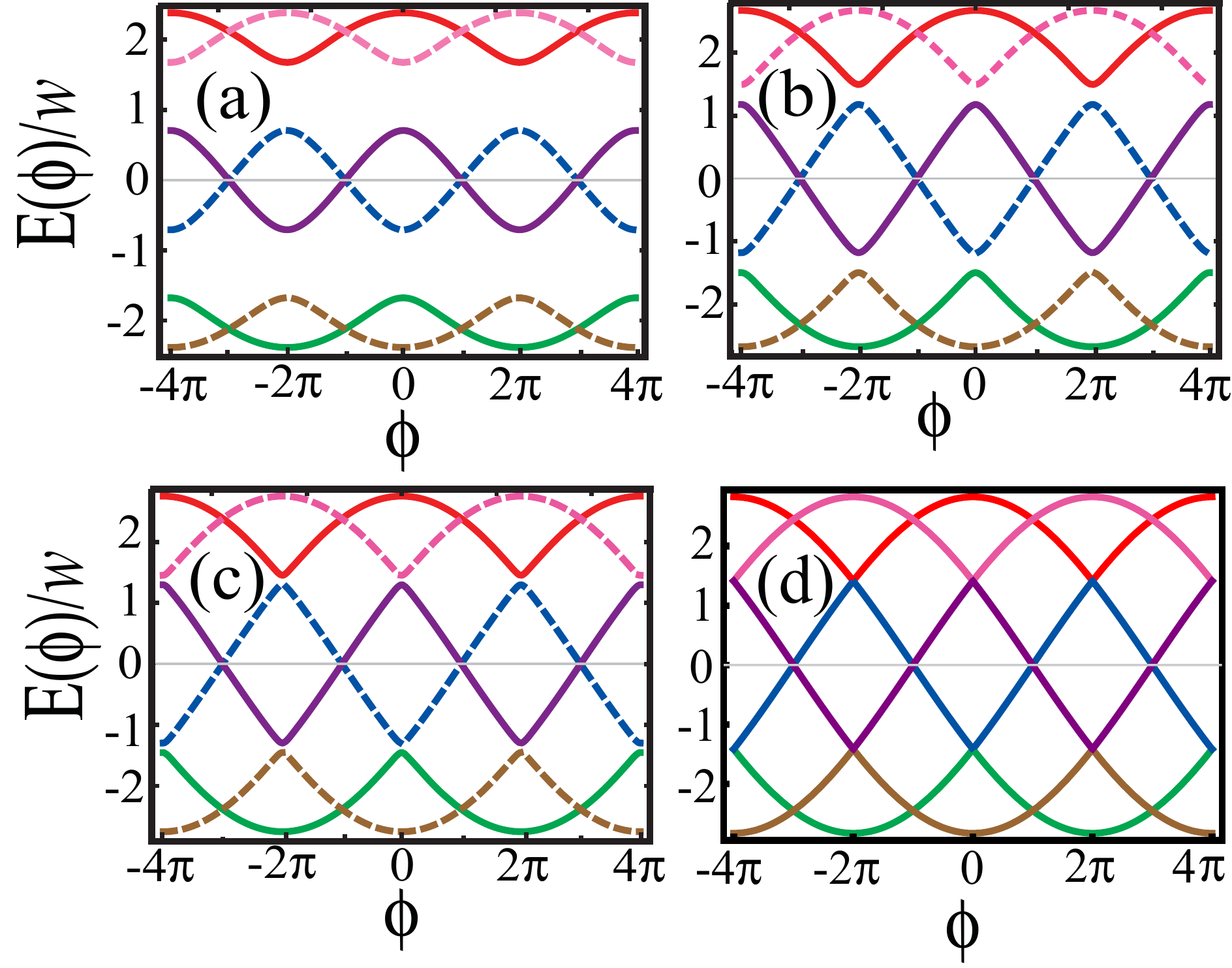}
\caption{Exact energy levels $E_{m\sigma}(\phi)/w$
for four different values of $g/w$: (a) $g/w=\sqrt{2}/2$; (b)
$g/w=\sqrt{2}/1.2$; (c) $g/w=\sqrt{2}/1.09$;
(d) $g/w=\sqrt{2}$. The color scheme of the curves here refer to the following.
The red and magenta curves are $ E_{+1,\sigma}(\phi)$, the green and brown ones
are
$E_{-1,\sigma}(\phi)$, while the blue and the purple ones represent
$E_{0,\sigma}(\phi)$. It is readily seen that curves $E_{m\sigma}(\phi)$ for a
given $m$ cross,
with a gap between the curves having different values of $m$. In Panel (d) the levels merge, being in this way equivalent to the energy curves
given in Eq. (\ref{e-crit}).}
\label{Fig:energies}
\end{center}
\end{figure}

In Fig. \ref{Fig:energies} we plot the energies (\ref{energies}) for four different
values of the ratio $g/w$, from weak- to strong-coupling. Note the
crossing between the levels $E_{m\sigma}(\phi)$ for a given $m$, with a gap between levels having different values of $m$.
In panel (a) of Fig. \ref{Fig:energies} we have a typical weak-coupling situation, which may be well described by the approximate formulas,
Eqs. (\ref{approx0}), (\ref{approx1}), and (\ref{approx2}). However, as the strong-coupled regime is approached, the gap between the levels having
different values of $m$ starts to close [panels (c) and (d) in Fig. \ref{Fig:energies}]. For $g/w=\sqrt{2}$ [panel (d) in the Fig. \ref{Fig:energies}]
the gap closes completely,
corresponding to a merging of three energy levels with quantum numbers $m=-1,0,1$.
For $g>\sqrt{2}w$ the gap opens again and it starts to grow for increasing $g$. For $g<\sqrt{2}w$ and for $g>\sqrt{2}w$ the levels
$E_{m\sigma}$ for a given $m$ are doubly degenerated at $\phi=(2k+1)\pi$, $k\in\mathbb{Z}$, while
no degeneracies occur at $\phi=2k\pi$ [See, e.g., panels (a), (b), and (c) at Fig. \ref{Fig:energies}]. However, for $g=\sqrt{2}w$
double degeneracies also occur at $\phi=2k\pi$, with the difference that these degeneracies also represent points of non-analyticity of
the energy eigenvalues as functions of $\phi$.

The level crossings at $\phi=\pi$ are particularly important, especially the ones crossing zero energy. Indeed, the level 
crossings at zero energy are associated to additional Majorana zero energy modes which are a linear combination of 
fermionic operators living on both sides of the junction. Thus, these additional Majorana modes arising for $\phi=\pi$ are 
very different from the ones living at the outer boundaries (the left end of chain A and the right end of chain B), 
$\gamma_1$ and $\delta_N$, which are independent of $\phi$. 
The zero energy Majorana modes at $\phi=\pi$ are given by
\begin{equation}
 \tilde \gamma_1=\frac{r}{\sqrt{2r^2+8}}\left[\frac{2i}{r}(a_N-a_N^\dagger)+b_2+b_2^\dagger\right],
\end{equation}
\begin{equation}
 \tilde \gamma_2=\sqrt{\frac{2}{r^2+4}}\left[\frac{ir}{2}(a_{N-1}-a_{N-1}^\dagger)+b_1+b_1^\dagger\right],
\end{equation}
where $r\equiv g/w$. The Majorana fermion $\tilde \gamma_1$ is a superposition of the Majorana fermions 
$\gamma_N=(i/\sqrt{2})(a_N^\dagger-a_N)$ and $\delta_2=(b_2+b_2^\dagger)/\sqrt{2}$, while $\tilde \gamma_2$ is 
given as a superposition of the Majorana fermions $\gamma_{N-1}=(i/\sqrt{2})(a_{N-1}^\dagger-a_{N-1})$ and 
$\delta_1=(b_1+b_1^\dagger)/\sqrt{2}$. Thus, for $\phi=\pi$ the total number of Majorana states (four, 
including the outer boundaries) is the same as the number of Majorana modes for two decoupled chains. Note that the decoupled chains limit 
$r\to 0$ yields $\tilde \gamma_1=\gamma_N$ and $\tilde \gamma_2=\delta_1$, as expected. These Majorana states at $\phi=\pi$ are 
robust and continue to exist even at very large Josephson couplings, $r\to\infty$, in which case they become 
$\tilde \gamma_1=\delta_2$ and $\tilde \gamma_2=\gamma_{N-1}$. Since varying $r$ corresponds to changes on the bulk properties 
of the system, due to the fact that $|\Delta|=w$, the zero Majorana modes at $\phi=\pi$ are topologically protected. 
Furthermore, their number ${\cal M}=\tilde \gamma_1^2+\tilde \gamma_2^2=1$, leading to a parity $(-1)^{\cal M}=-1$ at 
the tunnel junction when $\phi=\pi$. By recalling Eq. (\ref{Eq:constraint-junction}), we see that for $m=0$ and $\phi=\pi$ 
the number operators $N_{0,+1}=\tilde \gamma_1^2=1/2$ and $N_{0,-1}=\tilde \gamma_2^2=1/2$. Therefore, the charge is 
fractionalized in this case.

\begin{figure}
\begin{center}
\hspace{-0.5cm} \includegraphics[width=12cm]{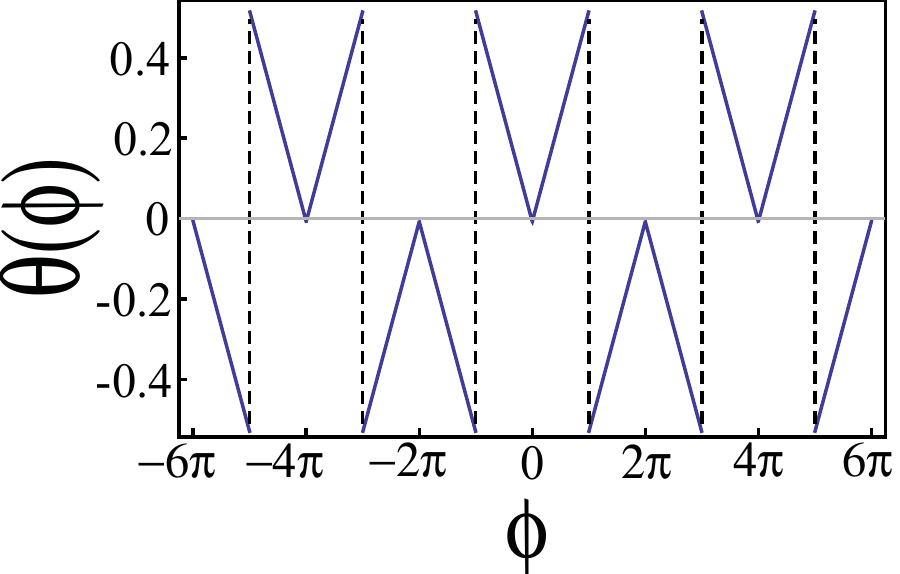}
\caption{Double-valued function $\theta(\phi)=(1/3)\arctan(|\sin(\phi/2)|/\cos(\phi/2))$ [Eq. (\ref{2-valued})].}
\label{Fig:theta}
\end{center}
\end{figure}

The additional level crossings at $g=\sqrt{2}w$ arise because for this value of $g$ the function $G(\phi)$ in Eq. (\ref{G}) becomes
$G(\phi)=|\sin(\phi/2)|$, which is non-analytic at $\phi=2 k\pi$. The energy levels become,
\begin{eqnarray}
\label{Ec}
 E_{m\sigma}(\phi)&=&\sqrt{2}w\left\{e^{-2\pi i(m-1)/3}\left[\sigma\cos\left(\frac{\phi}{2}\right)
-i\left|\sin\left(\frac{\phi}{2}\right)\right|\right]^{1/3}
\right.\nonumber\\
&+&\left.e^{2\pi i(m-1)/3}\left[\sigma\cos\left(\frac{\phi}{2}\right)
+i\left|\sin\left(\frac{\phi}{2}\right)\right|\right]^{1/3}\right\}.
\end{eqnarray}
If we consider $m=\sigma=1$ and $-2\pi\leq\phi\leq 2\pi$, Eq. (\ref{Ec}) simplifies to,
\begin{equation}
 E_{11}(\phi)=2\sqrt{2}w\cos(\phi/6).
\end{equation}
The muultivaluedness of $E_{m\sigma}(\phi)$ leads to
\begin{equation}
 E_{11}(\phi)=2\sqrt{2}w\cos\left(\frac{\phi}{6}-\frac{2\pi}{3}\right),
\end{equation}
in the interval $2\pi\leq\phi\leq 6\pi$, which is the same functional form as $E_{01}(\phi)$ in the
interval $0\leq\phi\leq 2\pi$.
By further analyzing the functional dependence of (\ref{Ec}) on $\phi$ and its quantum numbers, we obtain that
the spectrum for $g=\sqrt{2}w$ shown in panel (d) of Fig. \ref{Fig:energies} is indistinguishable from an energy spectrum
of the form,
\begin{equation}
\label{e-crit}
 \tilde E_{m\sigma}(\phi)=2\sigma\sqrt{2}w\cos\left[\frac{\phi}{6}+\frac{2\pi(m-1)}{3}\right].
\end{equation}
More precisely, the energy levels above are the principal values of the energies given in Eq. (\ref{Ec}). This can be seen from
Eq. (\ref{Eq:energies-2-valued}), which for $g=\sqrt{2}w$ becomes
\begin{equation}
\label{e-crit-true}
 E_{m\sigma}(\phi)=2\sigma\sqrt{2}w\cos\left[\theta(\phi)+\frac{2\pi(m-1)}{3}\right],
\end{equation}
where
\begin{equation}
\label{2-valued}
\theta(\phi)=\frac{1}{3}\arctan\left(\frac{|\sin(\phi/2)|}{\cos(\phi/2)}\right),
\end{equation}
is now a double-valued function of $\phi$; see Fig. \ref{Fig:theta}.
Eq. (\ref{e-crit-true}) describes the nature of the special point $g=\sqrt{2}w$
more accurately than Eq. (\ref{e-crit}), although both sets of functions lead to the curves shown in panel (d) of Fig. \ref{Fig:energies}.
Indeed, in Eq. (\ref{e-crit-true}) we can still see that the $4\pi$-periodicity of the spectrum via the double-valued
function (\ref{2-valued}), while this is not any longer apparent when the functions (\ref{e-crit}) are used. Note that
$\theta(\phi)$ is non-analytic at $\phi=2k\pi$. From this we immediately see that for $g=\sqrt{2}w$ the Josephson 
current will
exhibit jumps at  $\phi=2k\pi$. This expectation will be confirmed by explicit calculations in the next Section.

For $g=\sqrt{2}w$ a tunneling event from the state having quantum numbers $m=1$ and $\sigma=\pm 1$ into one having
$m=0$ and $\sigma=\pm 1$ is allowed at $\phi=2k\pi$. In fact, we
can see from panel (d) in Fig. \ref{Fig:energies} that there is for $g=\sqrt{2}w$ an eigenstate with
energy $E_{1,-1}(0)=E_{0,-1}(0)$, thus allowing a transition to the eigenstate having quantum numbers
$m=0$ and $\sigma=-1$, which crosses zero at $\phi=\pi$, corresponding to a Majorana zero mode there.  Furthermore, if we use the energies
(\ref{e-crit}), all curves cross zero for some value $\phi=(2k+1)\pi$.
A similar behavior to Eq. (\ref{e-crit-true}) is obtained in a three-junction Josephson
ring with a magnetic flux inside the triangular loop \cite{Nazarov-Book}. In this case $m$ would correspond to the number of vortices.
In the case of Eq. (\ref{e-crit-true}) each junction in the triangular loop would refer to a $Z_2$ fractionalized Josephson effect,
i.e., one featuring half of the phase difference.

\begin{figure}[ht]
\centering
\includegraphics[width=\linewidth]{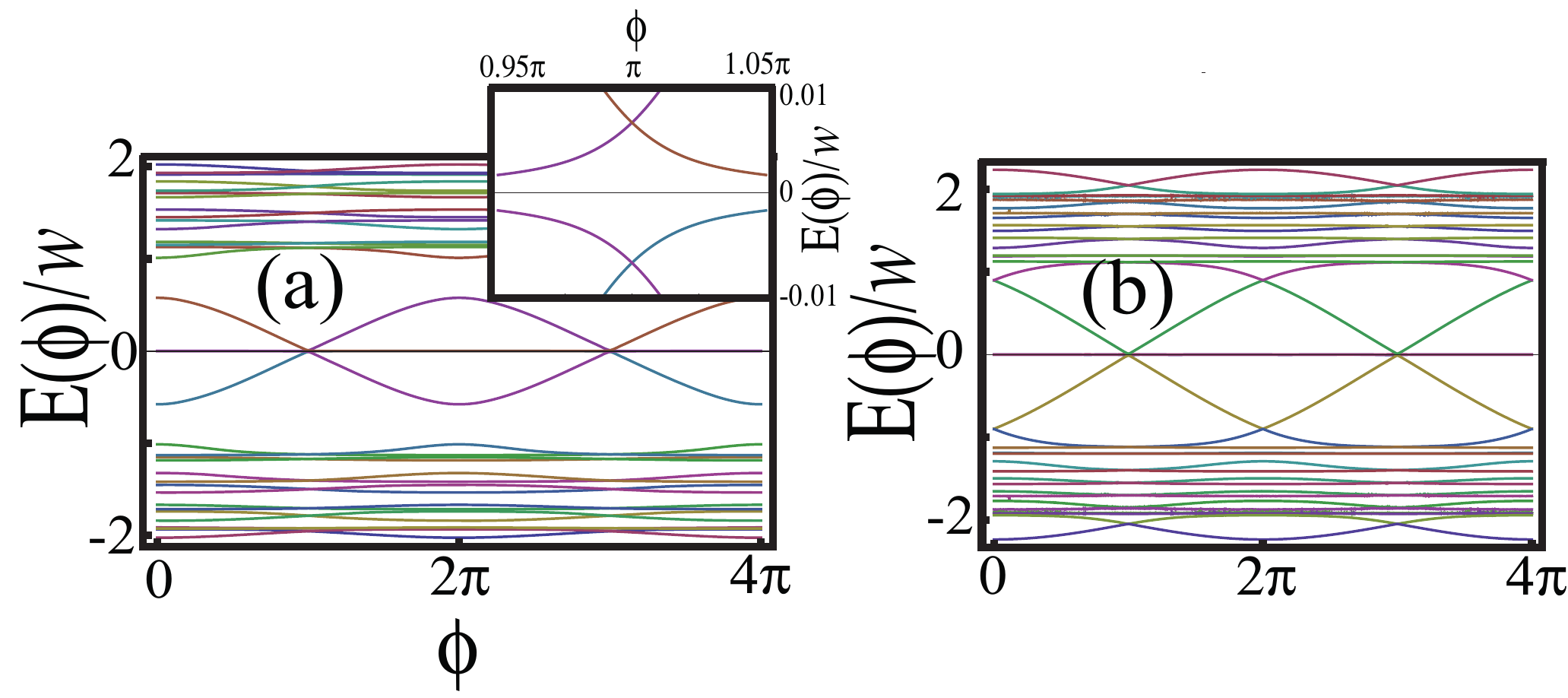}
\caption{Energy levels for the tunnel junction involving two Kitaev chains, each with $N=9$ (making a total
system size of 18 sites), and $|\Delta|=w/2$,
for (a) $g<g_{c}$ and (b) $g=g_c=\sqrt{5}w/2$. The inset in (a) zooms the region
around $\phi=\pi$ for $g<g_c$. Note that there is actually no exact crossing at $\phi=\pi$ and at zero energy, in contrast to
$|\Delta|=w$, which indicates a small overlap of Majorana fermions.
}
\label{fig:general}
\end{figure}

The result of Eq. (\ref{Ec}) is quite interesting as it does not follow from the perturbation
theory \cite{Alicea-2010} and was not anticipated in the original work \cite{Kitaev}. In order to see whether this degeneracy found for  $g=\sqrt{2}w$ is a general feature of the model we consider also
the general case where $|\Delta|\neq w$. Although in this case an exact solution is still possible for each
individual chain \cite{Kitaev,Lieb-1961}, to solve exactly the Josephson system involving the two chains
analytically is not an easy task. Observe that in this relevant case one has to take into account that chains must have free ends.
As a result, the momenta will not belong to
the first Brillouin zone, but will satisfy a transcendental equation \cite{Lieb-1961}. Furthermore,
zero boundary modes (Majorana modes) are only present
for odd $N$ and, strictly speaking, the crossings of the energy levels at zero energy for $\phi=(2k+1)\pi$
do not any longer occur if $N$ is not large enough. In Fig. \ref{fig:general} we show the results for the exact diagonalization of the Hamiltonian (1)-(3) for $|\Delta|=w/2$  and 18 sites ($N=9$ for each of the chains). Fortunately, the value of $N$ for which level crossing occurs in a similar way as in Fig. \ref{Fig:energies} is not too large and already for $N=9$ results similar to the exact solution can be be found. In Fig. \ref{fig:general}(a) we show the results of the diagonalization for
$g<g_c=\sqrt{5}/2$.  Observe that in contrast to the exact analytical results there is no actual crossing of levels at zero
energy for $\phi=\pi$ and $N=9$, due to a small, almost negligible overlap of the Majorana fermions; see the inset in
the Fig. \ref{fig:general}-(a). This is a finite-size effect and as $N$ further increases, this small gap at $\phi=\pi$
 becomes exponentially small, so that the crossings at
zero energy for $\phi=(2k+1)\pi$ from  Fig. \ref{Fig:energies} are approximately recovered. We have confirmed this behavior with system sizes up to $N=50$ per
chain.
Another interesting result is that for increasing $g$ to the critical value, which is here $g_c=\sqrt{5}w/2$, one finds additional crossings
at $\phi=2k\pi$ [Fig.\ref{fig:general}(b)], again in correspondence with the exact solution for the case where $|\Delta|=w$.
This confirms that the analytical results, obtained for $|\Delta| = w$, are quite generic and do not depend on the choice of the parameters.

\section{Topological Josephson current}

The standard way for calculating the Josephson current is given by the following formula (in unities of $2e/\hbar$),
\begin{equation}
\label{standard}
 I=\frac{\partial F(\phi)}{\partial\phi},
\end{equation}
where $F(\phi)$ is the free energy of the junction as a function of the phase difference $\phi$. In the above equation all
states are taken into account, since standard equilibrium thermodynamics is used, i.e., the partition function is
simply given by $Z={\rm Tr}[e^{-\beta(H-\mu N)}]$. However, in the case studied here the parity of the states involved plays an
important role. This point has been extensively discussed in the literature \cite{Fu-Kane-2009,Kitaev,Feigelman-2010,Law-Lee,vanHeck-2011}.
Superconducting tunneling in systems with fixed parity \cite{Tinkham} grew in importance in the 1990s in view of
experiments performed with small tunnel junctions \cite{Tuominen-1992}. 
In the case of topological superconductors the role of fermionic parity is
even more crucial, as the parity of states is associated to the presence or
absence of Majorana boundary modes \cite{Kitaev,vanHeck-2011}. 

\begin{figure}
\centering
\begin{tabular}{cc}
\includegraphics[width=8cm]{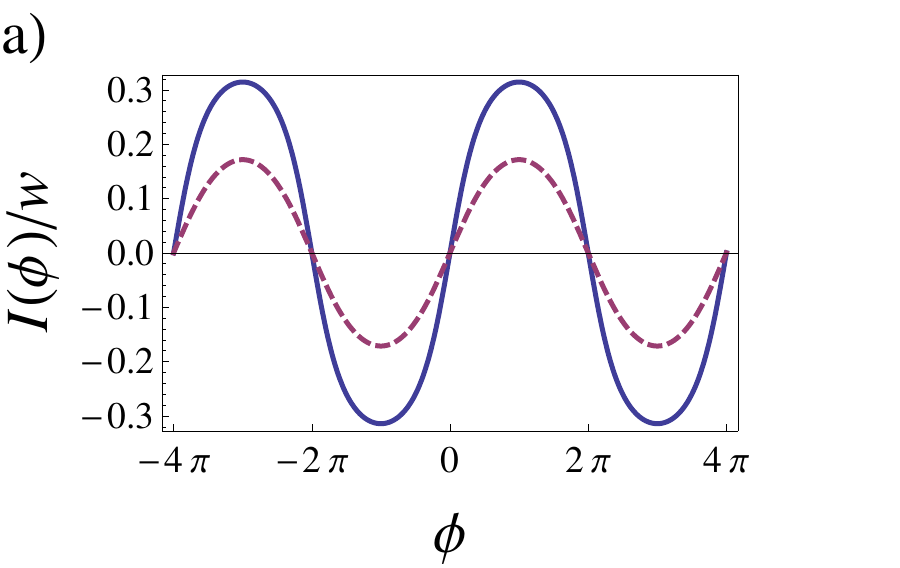} &
\includegraphics[width=8cm]{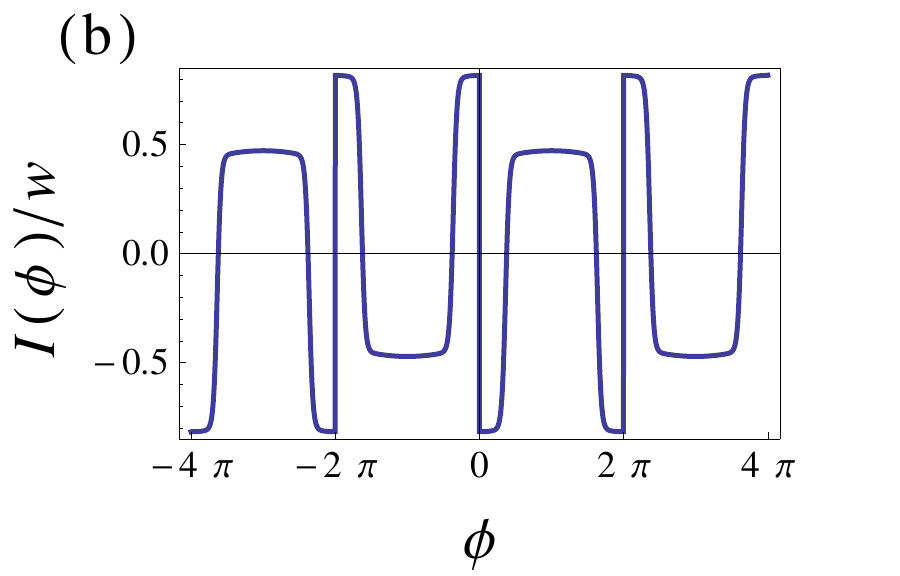}
\end{tabular}
\caption{Plots showing the Josephson current for fixed parity and low temperature ($T=w/10$). In Panel (a) we show
the Josephson current for two different values of $g$, namely, $g/w=\sqrt{2}/2$ (solid line) and $g/w=\sqrt{2}/4$ (dashed line).
Panel (b) shows the Josephson current for the critical value of the Josephson coupling, $g/w=\sqrt{2}$.}
\label{Fig:I-parity}
\end{figure}

The correct tunnel current is calculated by projecting out fixed parity states in the partition function. In order to do so,
we first note that only
three states are physically distinct. Indeed, as usual in any Bogoliubov treatment of superfluid systems, redundant states are introduced and
quasi-particles occur in pairs and the energy spectrum includes in this way the energies of particle and hole states \cite{Tinkham}.
In our case, this fact is expressed in relation (\ref{Eq:energ-parity}). 
Therefore,
we can compute the partition function for fixed parity using just three states out of the six ones
determined by Eq. (\ref{energies}), corresponding to the energy levels $E_1\equiv E_{0,-1}$, $E_2\equiv E_{1,+1}$, and
$E_3\equiv E_{1,-1}$. In other words, we simply have to use the effective Hamiltonian (\ref{Eq:H-Junction}).  
If
$N\equiv N_1+N_2+N_3$, the projection operators for even and odd parity states are given by \cite{Janko-1994},
\begin{equation}
 P_{e/o}=\frac{1}{2}[1\pm (-1)^N].
\end{equation}
The partition functions for fixed even and odd parities are
\begin{equation}
 Z_{e/o}={\rm Tr}[P_{e/o}e^{-\beta H_{\rm Junction}}],
\end{equation}
respectively. Thus, in $Z_e$ only even parity states are retained, with odd parity states being suppressed. The opposite
is true for $Z_o$, where even parity states are suppressed.
For the computation of the Josephson current we have to use the free energy \cite{Feigelman-2010},
\begin{equation}
 F(\phi)=\frac{1}{\beta}\ln\left(\frac{Z_e}{Z_o}\right).
\end{equation}
In Fig. \ref{Fig:I-parity} we show plots of the Josephson current in the case of fixed parity and low temperature.
The plots shown in
Fig. \ref{Fig:I-parity} are essentially zero temperature ones; there is no appreciable difference between the current profiles
shown and the ground state result.

Nevertheless the closure of the gap between the different energy levels at $g/w=\sqrt{2}$ remains manifest
in the calculations of the current at fixed parity. This is clearly seen from Fig.\ref{Fig:I-parity}(b) where discontinuous jumps
at $\phi=2\pi k$ occur due to additional degeneracies in the system.  Mathematically the origin of the discontinuities
comes from the fact that the function $G(\phi)$ in Eq. (\ref{G}) becomes $G(\phi)=|\sin(\phi/2)|$ for $g=\sqrt{2}w$, which
is non-analytic at the points $\phi_k=2k\pi$, making $dG/d\phi$ discontinuous at these points. We can interpret these jumps
physically in a way similar to the one of a so called SNS junction, where the energy of an Andreev bound state is given by
$E=\Delta\sqrt{1-{\cal T}^2\sin^2(\phi/2)}$ (we have considered for simplicity only one bound state) \cite{Beenakker}. In the
case of a fully transparent normal interface the transmission coefficient becomes ${\cal T}=1$, and $E=\Delta|\cos(\phi/2)|$. Thus,
inspired by this result, we can interpret the quantity,
\begin{equation}
 {\cal T}=\frac{2gw^2}{\left(\frac{g^2+4w^2}{3}\right)^{3/2}},
\end{equation}
as a transmission coefficient. For $g=\sqrt{2}w$ the junction becomes fully transparent.

\begin{figure}
\centering
\includegraphics[width=8cm]{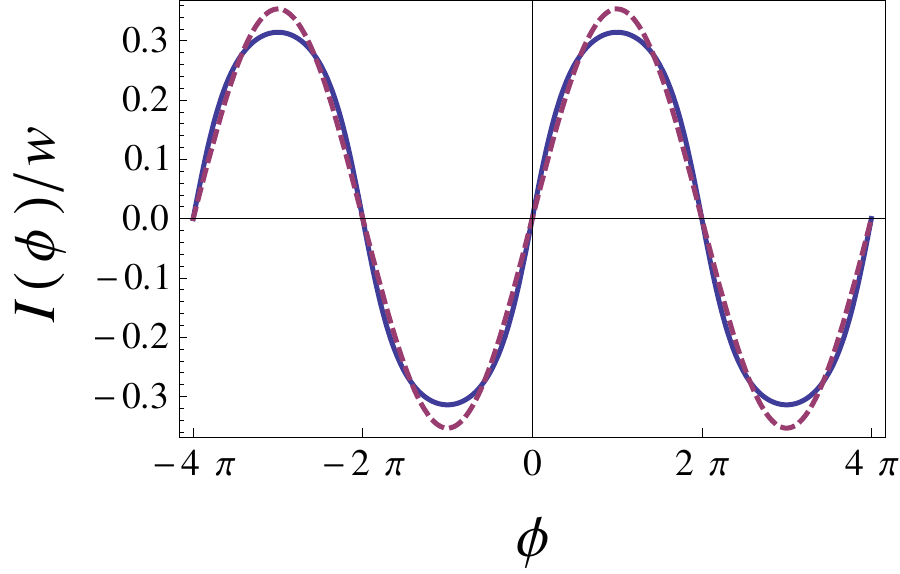}
\caption{Comparison between the exact result (solid line) and the second-order perturbative result \cite{Alicea-2010} (dashed line)
for $g=\sqrt{2}/2$ and $T/w=0.01$. In this example the perturbative result overestimate the maximum current by about 11 $\%$.}
\label{Fig:exactvspert}
\end{figure}

We also note in passing that the exact results at weak-coupling, although being qualitatively similar to the second-order perturbative result \cite{Alicea-2010}, still differ with respect to the the maximum value of
the current, as shown in Fig. \ref{Fig:exactvspert}. At the same time,
for the critical $g/w=\sqrt{2}$ the perturbative results look completely different as they miss
an additional feature associated with the closure of the energy gap.

It is well known that superconductivity cannot occur in one-dimensional systems at finite temperature. However, in the Kitaev model
the $U(1)$ symmetry is {\em explicitly} broken by proximity effect. We are not dealing with spontaneous symmetry breaking here, which
is actually a property of the higher dimensional substrate over which the wire is placed.
Indeed, the gap in the wire is held fixed, a situation
that may be approximately achieved using a proximity effect with a superconductor at higher dimensionality. The
wire can be assumed to be made of a semiconducting material with strong spin-orbit coupling, which eventually becomes
an one-dimensional $p$-wave superconductor via proximity effect with the surface of an $s$-wave superconductor \cite{Kitaev,Alicea-2010}.

\begin{figure}
\centering
\begin{tabular}{cc}
\includegraphics[width=8cm]{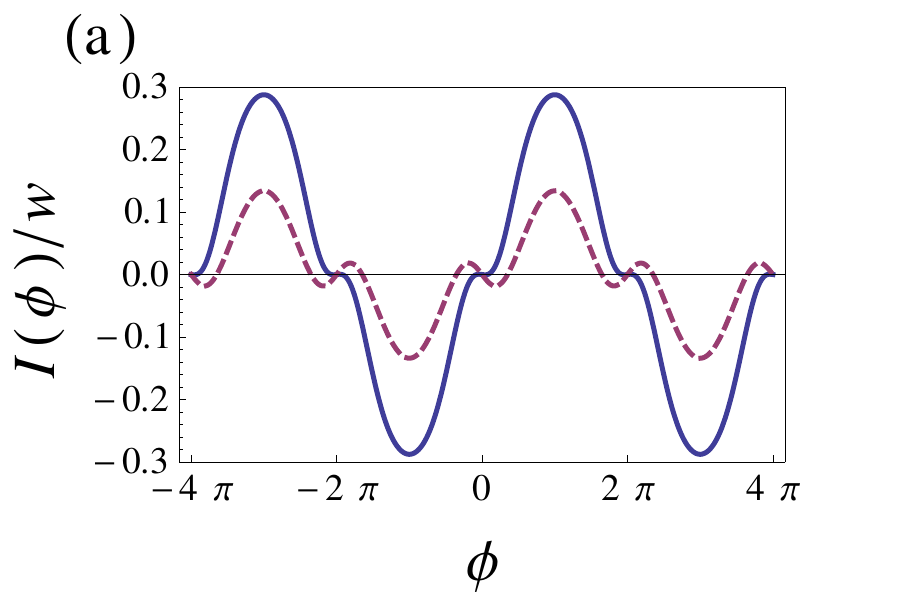} &
\includegraphics[width=8.15cm]{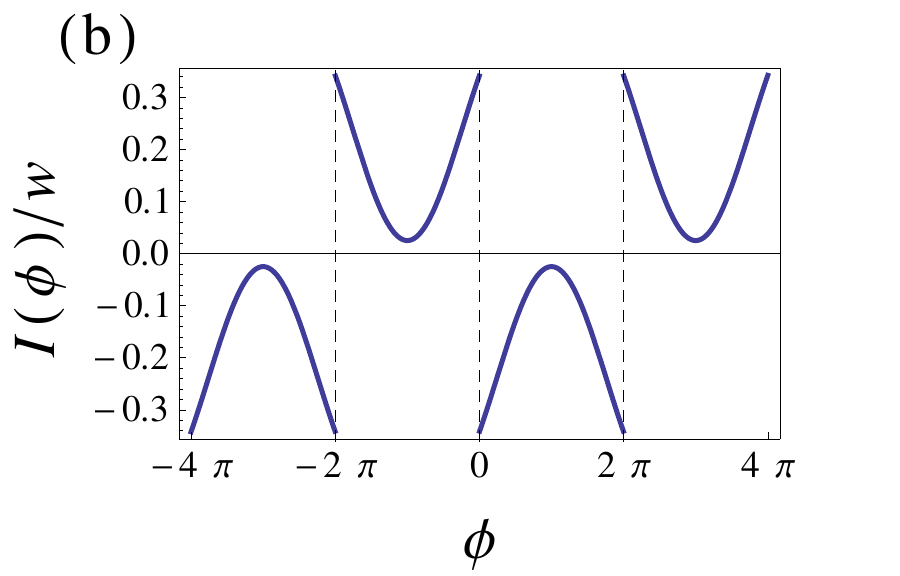}
\end{tabular}
\caption{Plots showing the Josephson current for fixed parity and high temperatures. In Panel (a) we show
the Josephson current for $g/w=\sqrt{2}/2$ and two different temperatures, namely, $T/w=2$ (solid line) and $T/w=1$ (dashed line).
Panel (b) shows the Josephson current for the critical value of the Josephson coupling, $g/w=\sqrt{2}$ and
$T/w=2$.}
\label{Fig:I-parity-T}
\end{figure}

In Fig. \ref{Fig:I-parity-T} we show plots of the Josephson current at finite temperatures. Not surprisingly, the
amplitude of the Josephson current decreases with the temperature [see Panel (a)], while the $4\pi$ periodicity
remains intact. Observe also that the Josephson current for $g/w=\sqrt{2}$ again shows a discontinuous behavior which points towards
the special character of the spectrum at this value of $g/w$. Observe that here the discontinuous jumps are 2$\pi$ modulated just
like in the low temperature case, but the $4\pi$ periodicity
remains intact as it should.



\section{Conclusion}

We have analyzed the Josephson effect for two lattice quantum wires
featuring fused
Majorana zero energy boundary modes at the tunnel junction.
In the weak-coupling regime the exact solution reproduces the perturbative result \cite{Alicea-2010} for the energy containing a
contribution $\sim \pm\cos(\phi/2)$ relative to the
tunneling of paired Majorana fermions. As the tunnel amplitude $g$ grows relative
to the hopping amplitude $w$, the gap between the energy levels gradually diminishes until it closes completely
at the critical value $g_c$, whose magnitude depends on the ratio of $|\Delta|/w$. For $|\Delta|/w=1$ and $g_c=\sqrt{2}w$, the Josephson energies can
be cast in the form
given by Eq. (\ref{Ec}), which is very different from the result obtained at weak-coupling.
Although this regime which occurs at $g > w$ is rather exotic from the point of view of its experimental realization,
it is still interesting to see that Kitaev's model is richer than it was originally anticipated. In addition,
the experimental setup for the realization of the Majorana fermions in quantum wires is still under discussion; see for example
Ref. \cite{Sarma-2010}.  Thus, it would be interesting to analyze whether an experimental setup allowing for large values of the coupling
$g$ between the chains can be engineered. Furthermore, such a system can in principle be engineered using ultracold fermions in
an one-dimensional optical lattice \cite{Bloch}. In this case a way to achieve the strongly coupled limit would be to consider
the distance between the chains as being smaller than the lattice spacing in the chains.
On the other side, it should be noted that the closure of the gap at $g_c$ may be spoiled in several ways in realistic systems, since
additional Andreev states occuring at larger Josephson couplings may lead to a small gap. Even in the context of the simple
fine-tuned model we solved here something different may occur if the quantum character of the phase difference, which is important
for sufficiently small systems, is taken into account. By this we mean to consider the phase of the order parameter as a quantum
operator conjugate  to particle number. In spite of the difficulties involving the definition of a hermitian phase operator
\cite{Nieto_1993}, a semi-classical
treatment is possible in Josephson nanosystems and the gap closing effect we have found may
disappear due to such a contribution \cite{Nazarov-Book}. However, in a setup where
superconductivity is induced via proximity effect,  the simple description in terms of the Kitaev model may apply, and
the new aspects we have discussed here is likely to be remain robust .
Another interesting question is whether
the additional degeneracies remain intact if interaction effects are included in the Kitaev Hamiltonian \cite{Loss-2011}, or if
disorder is present in the system. As far as the topological stability of the Majorana fermions at the boundaries is concerned,
recent work indicates that they are stable against disorder \cite{Beenakker-2011}. This does not necessarily mean that
our additional degeneracies and 1/6 fractionalization at $g=g_c$ survives disorder effects. This aspect of the problem needs
further investigation, being beyond the scope of the present study.

One further interesting aspect of the problem we have discussed concerns parity effects on the Josephson current. The
standard equilibrium calculation where the parity is allowed to change would completely fail to account for the
$4\pi$ periodicity of the topological Josephson effect. In the standard equilibrium calculation a sudden change of sign
would occur in the Josephson current when $\phi=(2k+1)\pi$, which in turn would spoil the $4\pi$ periodicity in
the current, despite the $4\pi$ periodicity of
the energies. On the other hand, when the calculation is done using a fixed parity ensemble \cite{Tuominen-1992},
the discontinuity jumps associated with the transition between the states with different parity disappear and
the $4\pi$ periodicity is restored. Nevertheless the closure of the gap between the different energy levels at $g/w=\sqrt{2}$
shows up via discontinuities in the current as a function of the phase.
but these discontinuities are not of the same type as the ones appearing in the calculations where the parity is allowed to change.

\section*{Acknowledgments}
The authors benefited from invaluable discussions with Felix von Oppen, Igor Karnaukhov, Yuli Nazarov, and Mikhail Fistul.
They would also like to thank Kostantin Efetov for pointing out the similarity of part of our results with
the topological properties of polyacetylene. We also thank Karl Bennemann for discussions
on several aspects of unconventional Josephson effects. We acknowledge support by the SFB Transregio 12 of the DFG.
FSN acknowledges also the financial support of the
Deutsche Forschungsgemeinschaft (DFG), grant KL 256/42-3.

\end{document}